\documentclass[twocolumn,aps,prl,superscriptaddress]{revtex4}

\usepackage{amsmath}
\usepackage{graphicx}
\usepackage{MnSymbol}

\DeclareFontFamily{U}{euc}{}%
\DeclareFontShape{U}{euc}{m}{n}{<-6>eurm5<6-8>eurm7<8->eurm10}{}%
\DeclareSymbolFont{AMSc}{U}{euc}{m}{n}
\DeclareMathSymbol{\umu}{\mathord}{AMSc}{"16}

\begin{document}

\title{Plastic deformation of rough metallic surfaces}

\author{ A. Tiwari}
\affiliation{PGI-1, FZ J\"ulich, Germany}
\affiliation{www.MultiscaleConsulting.com}
\author{ A. Almqvist}
\affiliation{Machine Elements, {Lule\aa} University of Technology, {Lule\aa} 97187, Sweden}
\author{ B. N. J. Persson}
\affiliation{PGI-1, FZ J\"ulich, Germany}
\affiliation{www.MultiscaleConsulting.com}

\begin{abstract}
We present experimental and theoretical results for the surface topography of a plastically deformed metallic (aluminum) block. When a hard spherical body (here a steel-, silica glass- or silicon nitride ball) with a smooth surface is indented in a metal block with a nominally flat, but still rough, surface, a spherical-cup-like indentation result due to plastic flow. The surface roughness in the indented region is, however, not entirely flattened. The long wavelength (macroasperity) content of the roughness result from the roughness on the original (aluminum) surface, but now plastically deformed. The roughness at short length scale, in the plastically deformed macroasperity contact regions, result from the roughness on the hard ball, and from inhomogeneous plastic flow.

We model the contact mechanics using the boundary element method, combined with a simple numerical procedure to take into account the plastic flow. The theory can semi-quantitatively describe the modification of the roughness by the plastic flow. Since the fluid leakage of metallic seals in most cases is determined by the long wavelength roughness, we predict that the leakage can be estimated based on the elastoplastic contact mechanics model employed here. 

The plastic deformations of surfaces of some glassy polymers is very different from what we observed for aluminum, which we attribute to strong work-hardening and to inhomogeneous plastic flow for the polymers. Thus the numerical procedure to account for the plastic flow proposed here cannot be applied to glassy polymers in general.
\end{abstract}

\maketitle


{\bf 1 Introduction}

The contact between metallic bodies occur in many applications, and often the contact pressure is so high as to generate plastic deformation, at least at the asperity level\cite{C1,C2,C3,C4,C5,C6,C7,C8,C9,C10}. In fact, because of surface roughness and the high elastic modulus of most metals, the contact pressure between asperities at short length scale can be very high even when the nominal contact pressure is low. Thus for metals in the area of real contact some plastic flow will almost always occur, at least during the first contact\cite{Tabor}.

Metallic seals are used in many applications involving very high fluid pressure differences, and in ultra high vacuum systems. Surface roughness and plastic flow highly affects leakage in metallic seals, since they are key factors in determining the surface separation in the non-contact area.
For elastic solids like rubber, contact mechanics theories have been developed for how to predict the fluid leakage rate, and it has been shown that they are in good agreement with experiments\cite{Lorenz1,Lorenz2}. The simplest approach assumes that the whole fluid pressure difference between the inside and outside of the sealed region, occur over the most narrow constrictions (denoted critical junctions), encountered along the largest open percolating non-contact flow channels.

For elastic solids numerical contact mechanics models\cite{Ref7}, such as the boundary element model, and the analytic theory of Persson\cite{BP,Alm}, can be used to calculate the surface separation at the critical junction and hence predict fluid leakage rates. For solids involving plastic flow, the surfaces will approach each other more closely than if only elastic deformations would occur. This will reduce the fluid leakage rate\cite{Per1,AA4}.

Here we will present the outcome of a study, where we experimentally explore the nature of the plastic deformation of the asperities of a sandblasted aluminum surface, but we believe the results should hold quite generally for other metals of interest such as steel, copper or bronze. We will also present results from numerical simulations of the experimental set-up, based on the boundary element method combined with a simple procedure to include plastic flow. More precisely, we employ the method presented in\cite{AA1} that assumes an elastoplastic model where a solid deforms elastically until the local pressure reaches a critical stress (the penetration hardness), after which it flows without strain hardening.
 
Recently, several studies of surface roughness and plastic flow have been reported using microscopic (atomistic) models\cite{Pas}, or models inspired by atomic scale phenomena that
control the nucleation and glide of the dislocations\cite{Nic3,Nic2,Nic1}. These models supply fundamental insight into the complex process of plastic flow, but are not easy to apply to practical systems involving inhomogeneous polycrystalline metals and alloys exhibiting surface roughness of many length scales. The approach we use here is less accurate but easy to implement, and it can be used to estimate the leakage rates of metallic seals. It remains, however, to test in detail how accurate the results are.

\vskip 0.3cm
{\bf 2 Experimental}

The aluminum block was indented with either a steel ball with $50 \ {\rm mm}$ diameter, or a silicon nitride ${\rm Si}_3 {\rm N}_4$ ball with $33.338 \ {\rm mm}$ diameter, or a borosilica glass ball with the diameter $30 \ {\rm mm}$. The normal (indentation) force was $40 \ {\rm kN}$. Indentation was done on a rectangular aluminum block with a polished surface, and on two sandblasted aluminum surfaces. The sandblasting was done with glass beads (spherical particles with smooth surfaces) of diameter $\approx 10  \ {\rm \umu m}$ for a time ranging from 5 to 8 minutes using 8 bar air pressure. The topography measurements were performed with Mitutoyo Portable Surface Roughness Measurement device, Surftest SJ-410 with a diamond tip with the radius of curvature $R = 1 \ {\rm \umu m}$, and with the tip--substrate repulsive force $F_{\rm N}=0.75 \ {\rm mN}$. The lateral tip speed was $v=50 \ {\rm \umu m/s}$.

From the the measured surface topography (line scans) $z=h(x)$ we calculated the one-dimensional (1D) surface roughness power spectra defined by
$$C_{\rm 1D} (q) = {1\over 2 \pi} \int dx \ \langle h(x) h(0) \rangle e^{i q x} \eqno(1)$$
where $\langle .. \rangle$ stands for ensemble averaging. For surfaces with isotropic roughness the 2D power spectrum $C(q)$ can be obtained directly from $C_{\rm 1D} (q)$  as described elsewhere\cite{Nyak,CarbLor}. For randomly rough surfaces, all the (ensemble averaged) information about the surface is contained in  the power spectrum $C(q)$. For this reason the only information about the surface roughness which enter in contact mechanics theories (with or without adhesion) is the function $C(q)$. Thus, the  (ensemble averaged) area of real contact, the interfacial stress distribution and the distribution of interfacial separations, are all determined by $C(q)$\cite{BP,Alm}. Note that, the moments of the power spectrum determines the often quoted standard quantities, which are output of most stylus instruments. Thus, for example, the mean-square (ms) roughness amplitude $\langle h^2 \rangle$ and the ms slope $\langle (dh/dx)^2\rangle$ are given by 
$$\langle h^2 \rangle = 2 \int_0^\infty dq \ C_{\rm 1D}(q),$$ 
and
$$\langle (dh/dx)^2 \rangle = 2 \int_0^\infty dq \ q^2 C_{\rm 1D}(q),$$
respectively.

Since the surface topography was measured only along line scans $z=h(x)$, for the numerical
contact mechanics simulations (see Sec. 4) we produced randomly rough surfaces using the random-phase-method described in Appendix D in Ref. \cite{R1}. For generating these surfaces we used the 2D surface roughness power spectra obtained as described above.

\begin{figure}
\includegraphics[width=0.95\columnwidth]{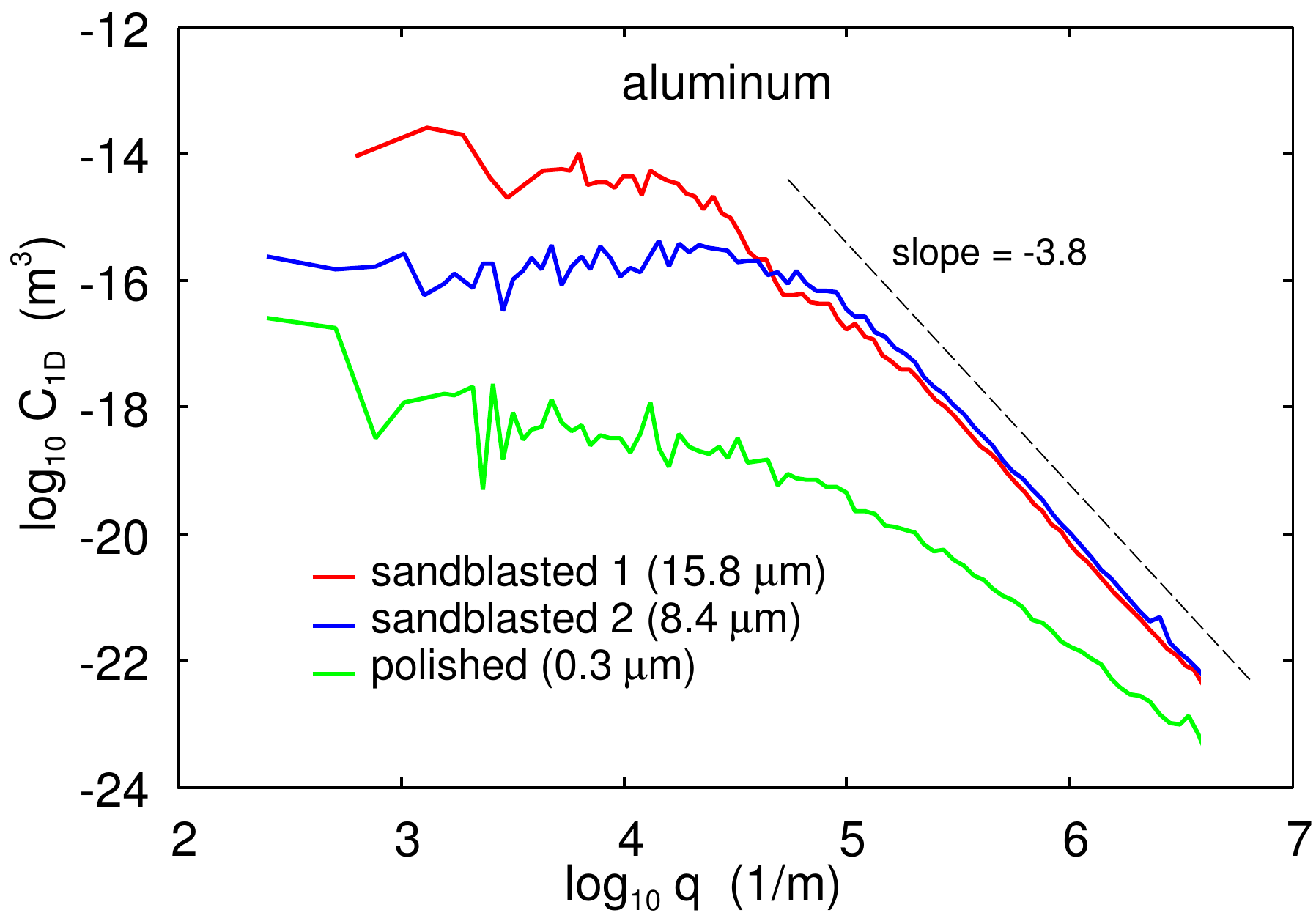}
\caption{\label{1logq.2logC1D.polished.sanblasted1.sandblasted2.pdf}
The surface roughness power spectrum as a function of the wavenumber (log-log scale)
for two sandblasted aluminum surfaces and one polished aluminum surface.
The root-mean-square (rms) roughness amplitude of the three surfaces are $h_{\rm rms} = 15.8$, $5.8$ and
$0.3 \ {\rm \mu m}$. The corresponding rms slopes are $0.42$, $0.45$ and $0.04$, respectively.
The surfaces have nearly vanishing skewness and the kurtosis is close to 3 for all the surfaces.
}
\end{figure}

\begin{figure}
\includegraphics[width=0.95\columnwidth]{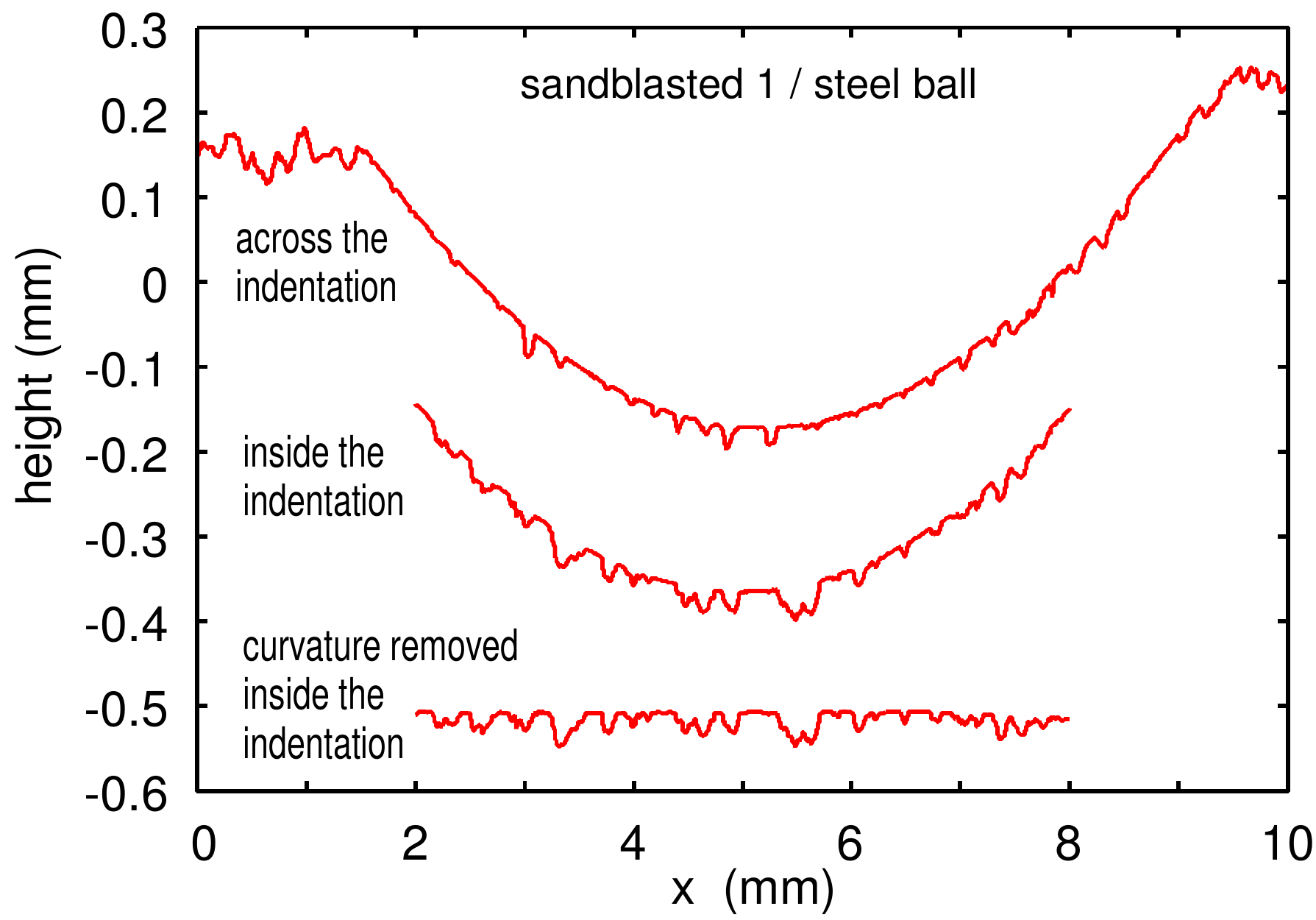}
\caption{\label{8minperptiltRemoved.pdf}
The surface roughness height profile $h(x)$ of the sandblasted aluminum surface 1
after squeezing a steel ball (radius $R=2 \ {\rm cm }$)
against the aluminum block with the axial force $F_{\rm N} = 40 \ {\rm kN}$ for 1 minute.
This result in a spherical cup indentation with the same radius of curvature as the steel ball
and with the indentation diameter $\approx 0.8 \ {\rm cm}$. The middle line shows the line scan data
from inside the indentation, and the bottom line after removing the macroscopic curvature. Note that the high asperities
have flat upper surfaces because of plastic flow.
}
\end{figure}

\begin{figure}
\includegraphics[width=0.95\columnwidth]{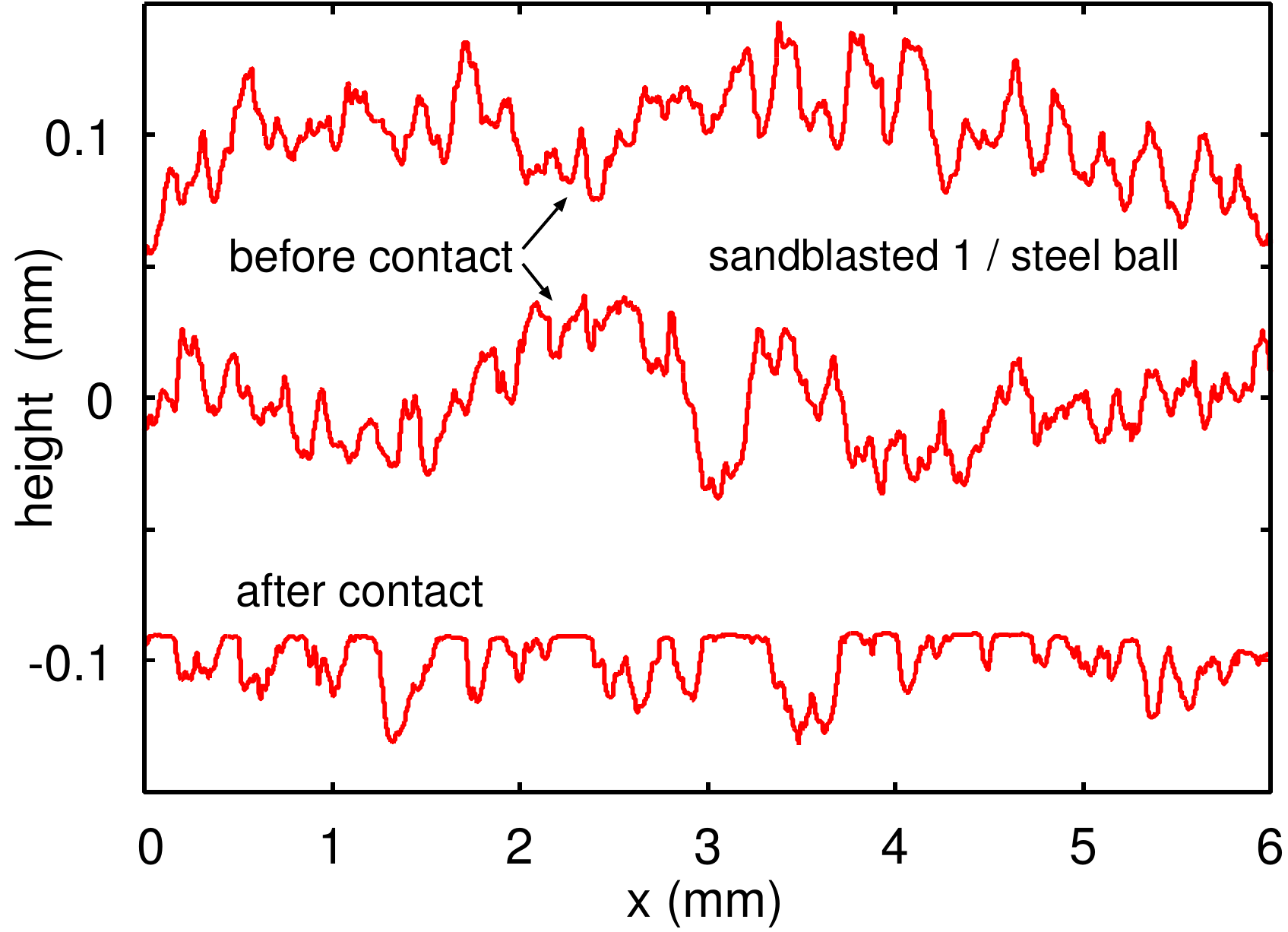}
\caption{\label{8minaluminRemoveCurvatureBeforeandAfterDeformation.pdf}
The surface roughness height profile $h(x)$ of the sandblasted aluminum surface 1
(top two lines), and after squeezing the steel ball
against the aluminium surface (bottom line).
The bottom linescan data is from inside the indentation by first removing the
macroscopic curvature. Note that the high asperities
have flat upper surfaces because of plastic flow.
}
\end{figure}

\begin{figure}
\includegraphics[width=0.95\columnwidth]{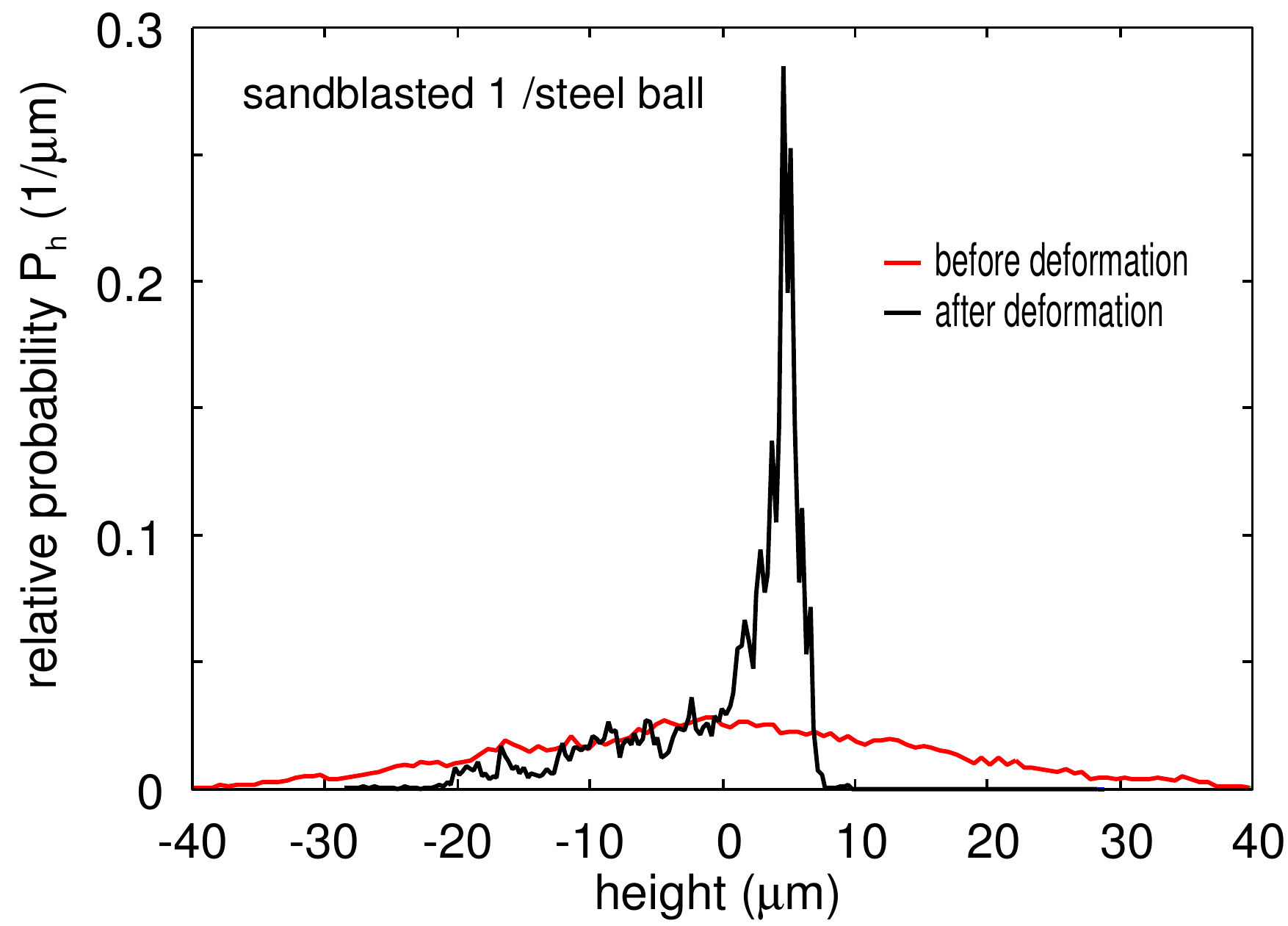}
\caption{\label{heightDistributionaluminBeforeAfterdeformation.pdf}
The surface roughness height distribution $P_h$ of the sandblasted aluminum surface 1
(red line), and after squeezing the steel ball
against the aluminum block (black line).
The black line is obtained from the line scan data
inside the indentation by first removing the macroscopic curvature. The sharp peak is due to the high asperities
have flat upper surfaces (of equal height) because of plastic flow.
}
\end{figure}

\begin{figure}
\includegraphics[width=0.95\columnwidth]{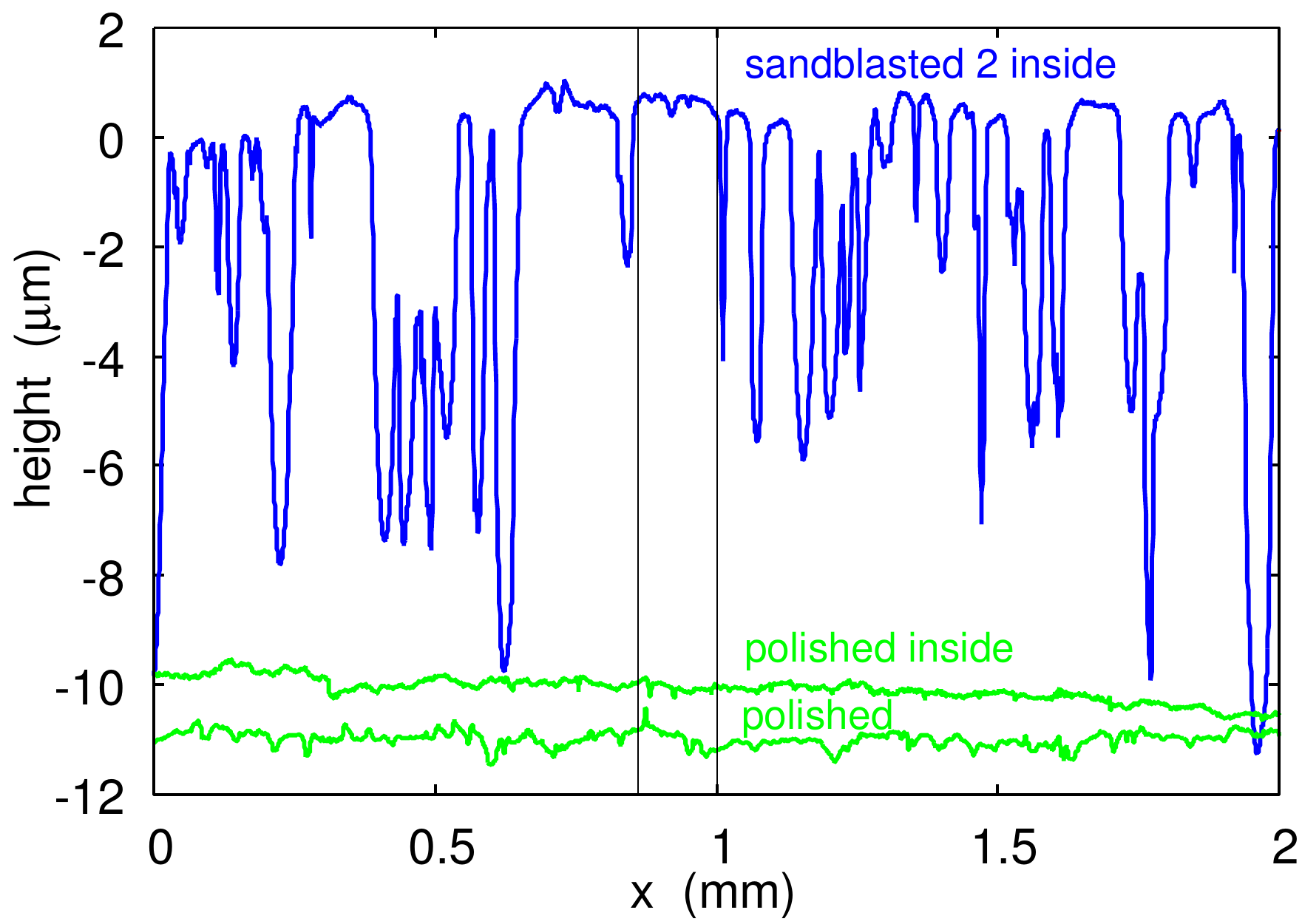}
\caption{\label{1x.2height.37.65.63.full.pdf}
The surface topography of the sandblasted aluminum surface 2 (top blue line), and of the polished
surface, indented with a ceramic ball
(Silicon nitride, ${\rm Si_3 N_4}$) with diameter 33.3 mm. The axial force $F_{\rm N} = 40 \ {\rm kN}$ for 1 minute.
The topography (middle line) is after removing the surface curvature. Also shown is the surface topography of the not indented
surface area of the polished ball (bottom green line).
}
\end{figure}

\begin{figure}
\includegraphics[width=0.95\columnwidth]{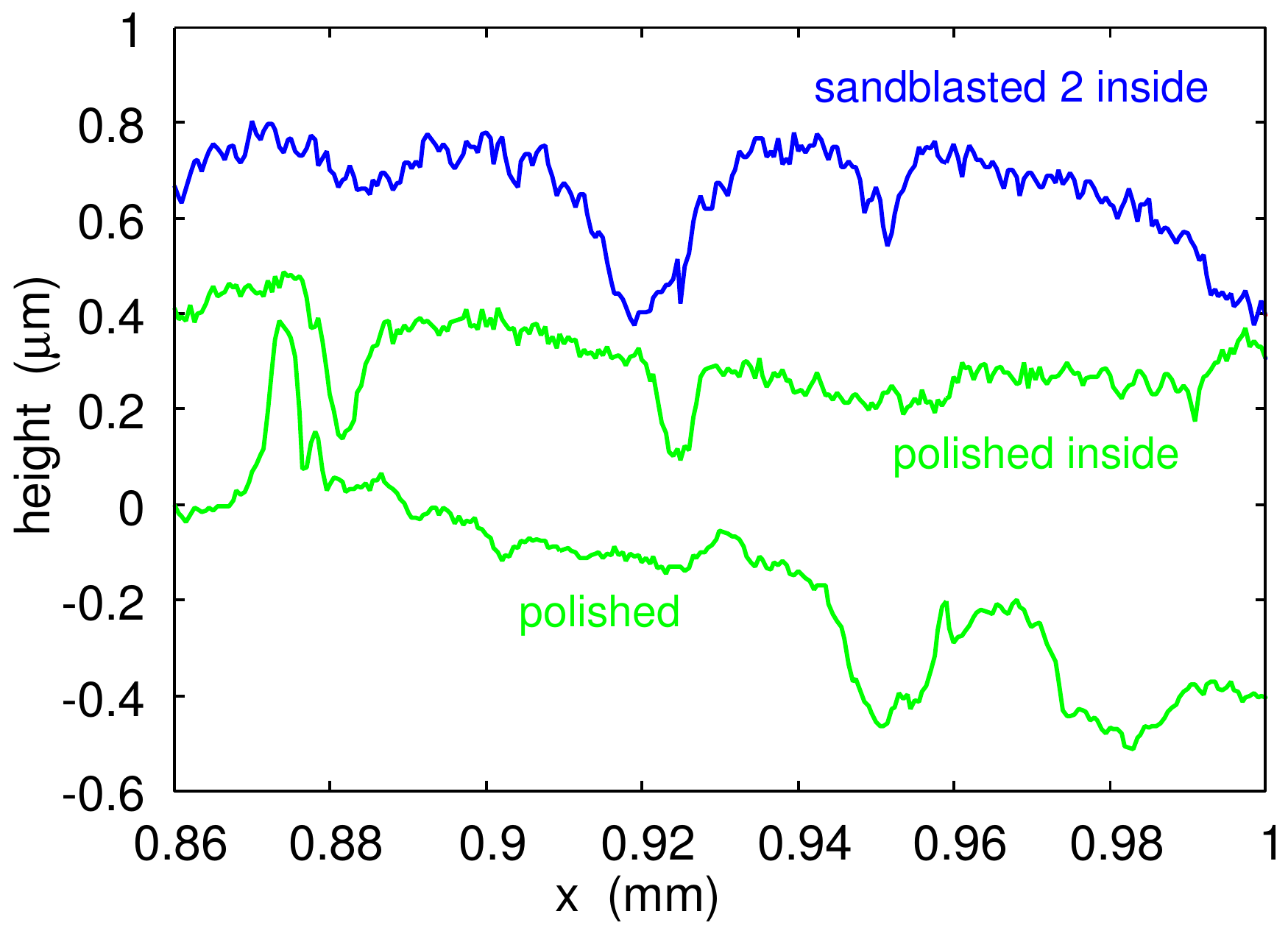}
\caption{\label{1x.2height.37.65.63.magnification.pdf}
The surface topography in Fig.\ref{1x.2height.37.65.63.full.pdf} between the two vertical
lines.
}
\end{figure}

\begin{figure}
\includegraphics[width=0.95\columnwidth]{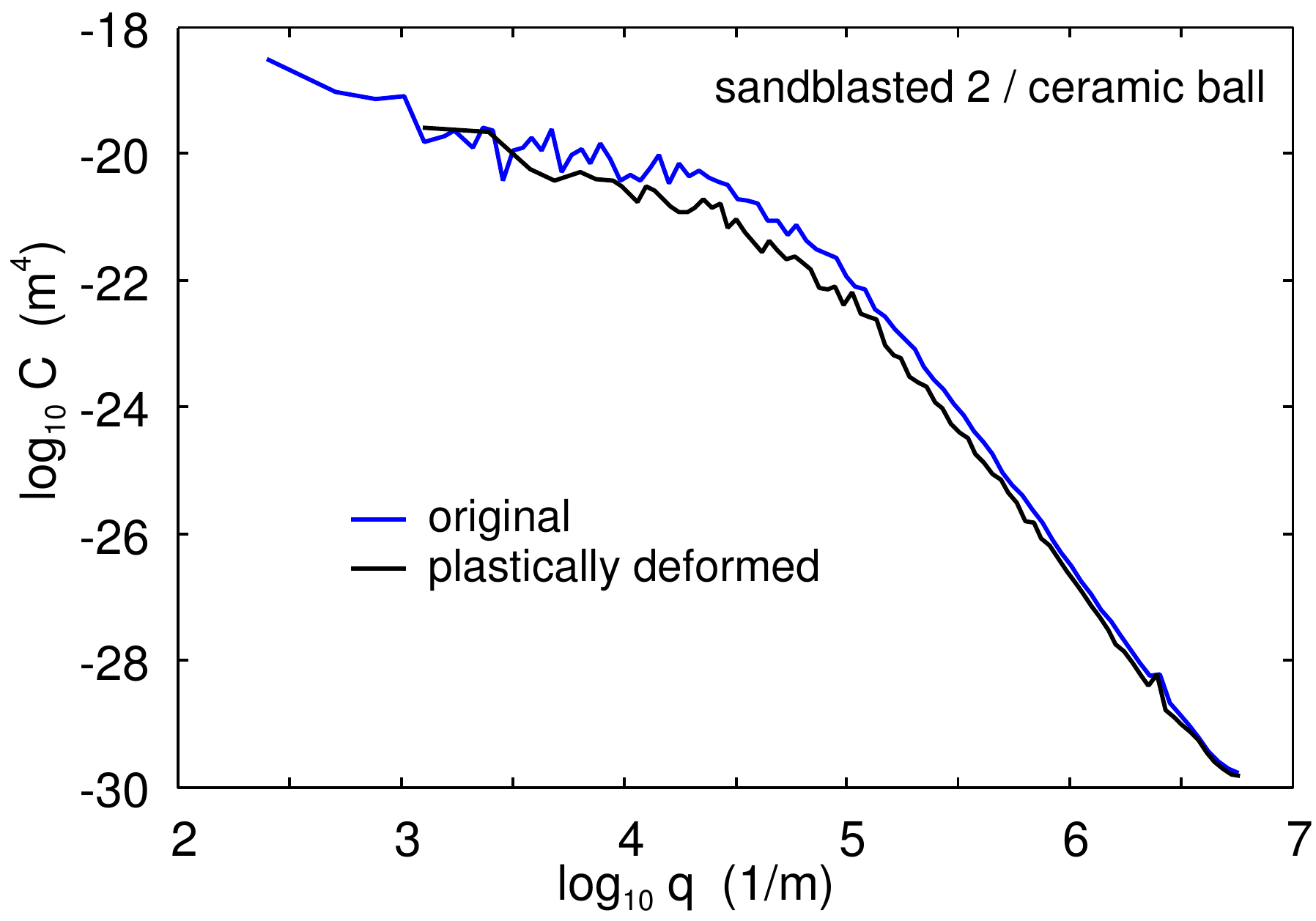}
\caption{\label{1logq.2logC.rough2.23.and.35-38.pdf}
The surface roughness power spectrum as a function of the wavenumber (log-log scale)
before indentation (blue), and after (black) plastic deformation.
For the sandblasted surface 2 indented by the ceramic ball.
}
\end{figure}

\begin{figure}
\includegraphics[width=0.95\columnwidth]{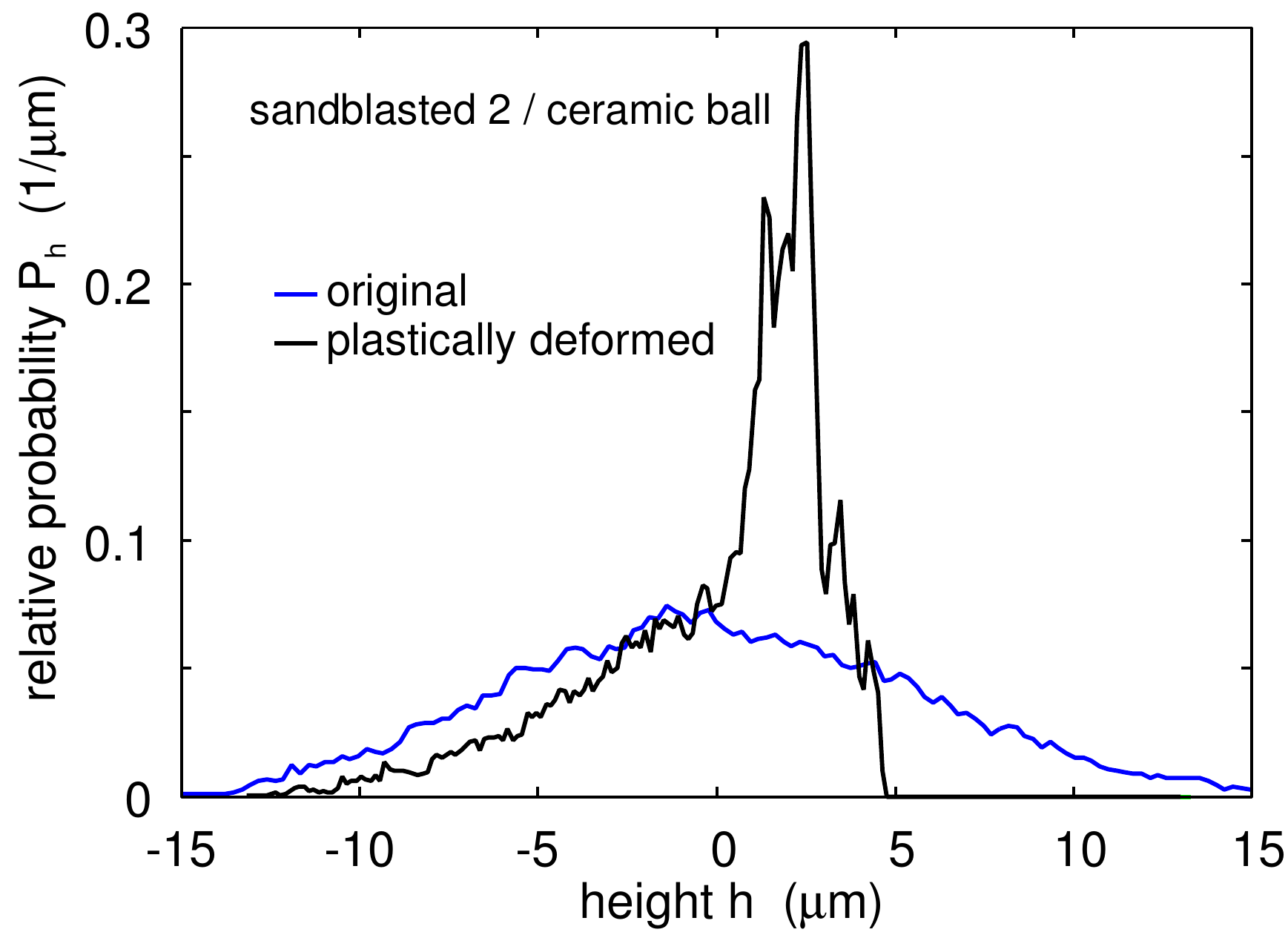}
\caption{\label{1h.2Ph.37.29.pdf}
The height probability distribution of the sandblasted surface 2
(blue line), and after indenting it with the ceramic ball (black line).
The height probability distribution of the indented surface
is from inside the indented region after removing the surface curvature.
}
\end{figure}

\begin{figure}
\includegraphics[width=0.95\columnwidth]{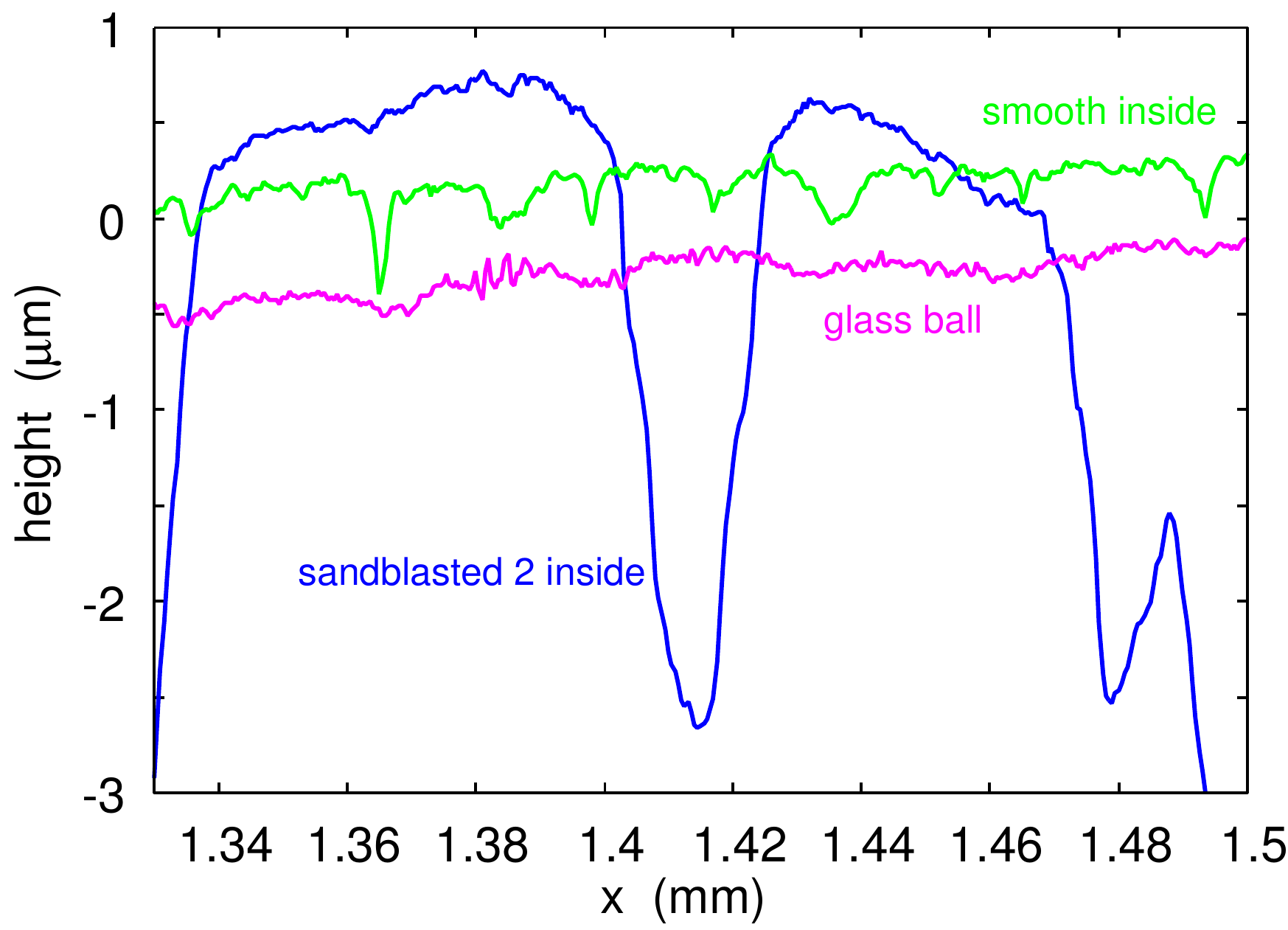}
\caption{\label{1x.2h.with.glass.ball.pdf}
The surface roughness height profile $h(x)$ of the sandblasted aluminum surface 2
after squeezing a glass ball (radius $R=1.5 \ {\rm cm }$)
against the aluminum block with the axial force $F_{\rm N} = 40 \ {\rm kN}$ for 1 minute (blue line).
This result in a spherical cup indentation with nearly the same radius of curvature as the glass ball.
The green line is the same result for a
polished aluminum surface. The pink line is the topography of the glass ball.
In all cases the curvature of the indentation (or the glass ball) is removed.
}
\end{figure}

\begin{figure}
\includegraphics[width=0.95\columnwidth]{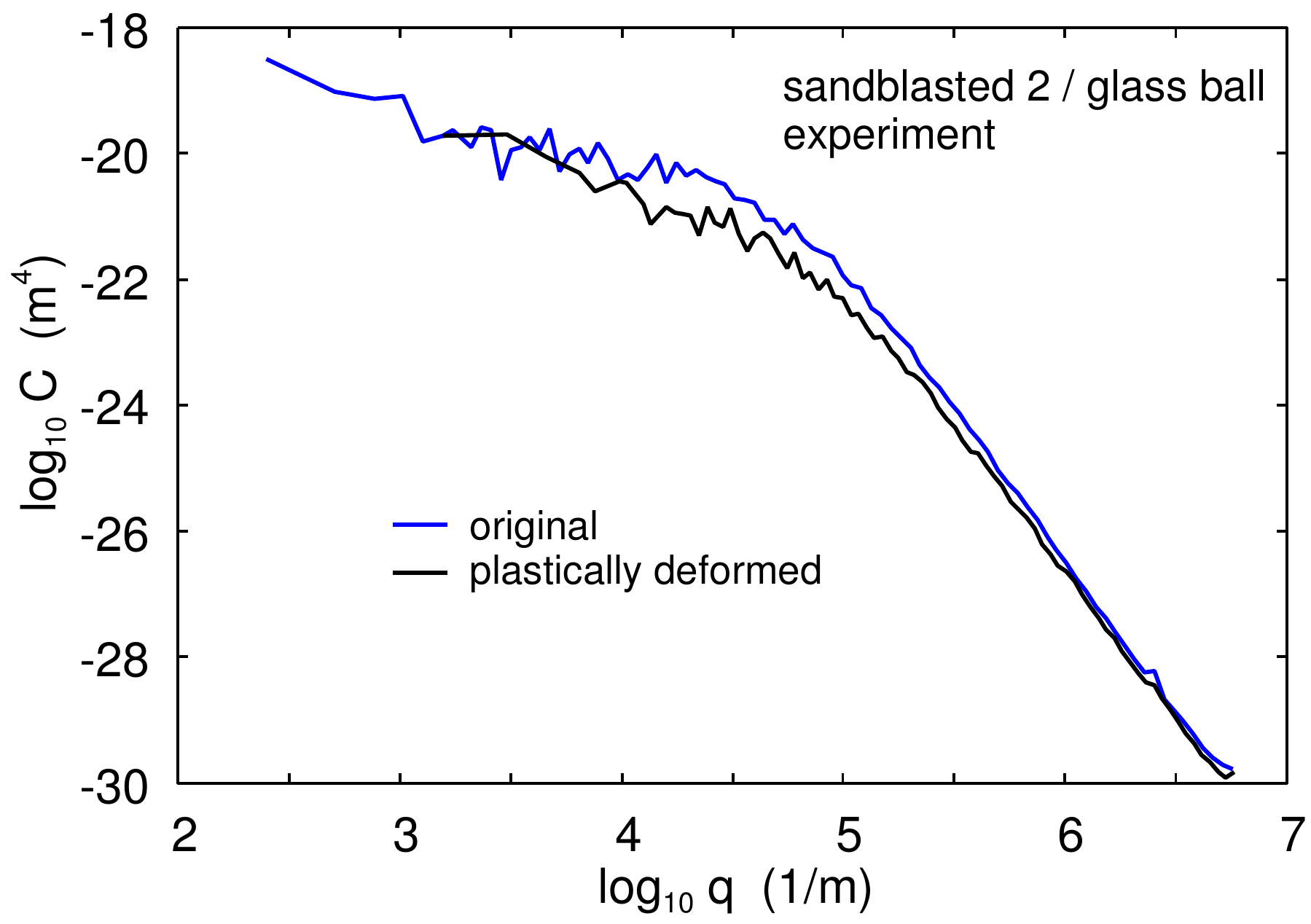}
\caption{\label{1logq.2logC.glassBall.sandblasted2.full.pdf}
The surface roughness power spectrum as a function of the wavenumber (log-log scale)
before indentation (blue), and after (black) plastic deformation.
For the sandblasted surface 2 indented by the glass ball and after removing the curvature of the indentation.
}
\end{figure}

\begin{figure}
\includegraphics[width=0.95\columnwidth]{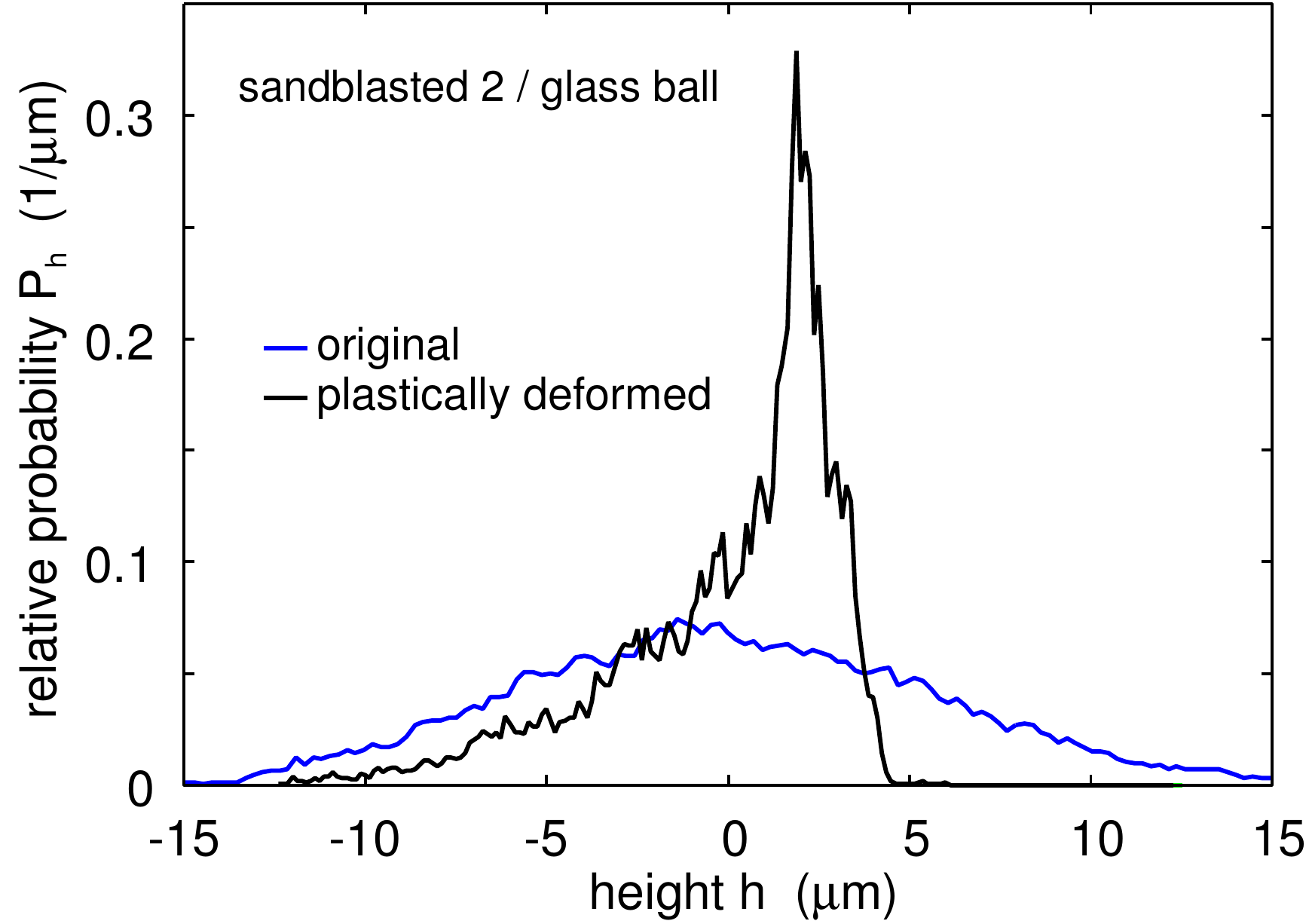}
\caption{\label{1h.2Ph.sandblasted2.glassBall.pdf}
The height probability distribution of the sandblasted surface 2
(blue line), and after indenting it with the glass ball (black curve).
The height probability distribution of the indented surface
is from inside the indented region after removing the surface curvature.
}
\end{figure}

\vskip 0.3cm
{\bf 3 Experimental results}

We have performed indentation experiments for three different nominally flat aluminum surfaces, one polished and two sandblasted. Before performing the indentation experiments we measured the surface  topography $z=h(x)$ along several $10 \ {\rm mm}$ long tracks, from which we have calculated the 1D surface roughness power spectra using (1).

Figure~\ref{1logq.2logC1D.polished.sanblasted1.sandblasted2.pdf} 
shows the surface roughness power spectra as a function of the wavenumber (log-log scale) for the sandblasted and polished aluminum surfaces. The root-mean-square (rms) roughness amplitude of the three surfaces are $h_{\rm rms} = 15.8$, $5.8$ and $0.3 \ {\rm \umu m}$, and the corresponding rms slopes are $0.42$, $0.45$ and $0.04$, respectively. The surfaces have nearly vanishing skewness ($-0.02$, $0.13$ and $-0.05$), and the kurtosis ($2.7$, $2.9$ and $3.1$) is close to 3 for all the surfaces, as expected for randomly rough surfaces with a Gaussian height probability distribution.

The sandblasted surface 1 was squeezed against a smooth steel ball, and the sandblasted surface 2 and the polished surface was squeezed against the ceramic ball and the silica glass ball, in all cases with the normal force $40 \ {\rm kN}$ for 1 minute. 

Figure~\ref{8minperptiltRemoved.pdf} 
shows the surface roughness height profile $h(x)$ of the sandblasted aluminum surface 1 after squeezing it against the steel ball. The spherical cup indentation has nearly the same radius of curvature as the steel ball, and with the indentation diameter $\approx0.8 \ {\rm cm}$. Also shown in the figure is a line scan from inside the indentation, before and after removing the macroscopic curvature. Note that the high asperities have flat upper surfaces because of plastic flow, while the roughness in the big valleys are left almost unchanged. This is very different from what we observed in Ref. \cite{Av} for glassy polymers, where during plastic deformation material moved effectively from the top of asperities to the nearby valley, resulting in  long wavelength roughness which appeared the same as on the original (undeformed) surface but with smaller amplitude. We attributed this to strong work-hardening. The aluminum surface is most likely already work-hardened by the production procedure, and this could explain the very different nature of the plastic flow in the two cases. 

Usually the material penetration hardness is defined as the ratio between the external normal force (here $F_{\rm N} = 40 \ {\rm kN}$) and the projected indentation area (here $A_0 = \pi r_0^2$ with $r_0 \approx 4 \ {\rm mm}$). This gives $\sigma_{\rm P} \approx 0.8 \ {\rm GPa}$. However, note that only $\approx 50\%$ of the nominal contact area in Fig.~\ref{8minperptiltRemoved.pdf} appears to be plastically deformed. In fact, if a plastically deformed macroasperity contact area is observed at higher magnification, then
short wavelength roughness can be observed, and an even smaller fraction than $\approx 50\%$ of the nominally contact area 
is plastically deformed (see also discussion below). We conclude that most likely the penetration hardness depends on the length scale (or size of the indentor), a fact which is well known from earlier studies using different size of the indentor, or different indentation depth\cite{Hard}. There are several different reason for this length-scale dependent hardness, e.g., it may result from a thin work-hardened surface layer. 


Figure~\ref{8minaluminRemoveCurvatureBeforeandAfterDeformation.pdf} 
shows magnified pictures of the surface topography in Fig.~\ref{8minperptiltRemoved.pdf}. The two upper curves are linescans from the sandblasted surface 1 before plastic deformation, and the lower curve shows a linescan from inside the indented region by first removing the macroscopic curvature. Note that the high asperities (at this magnification) appear to have flat upper surfaces because of plastic flow.

Figure~\ref{heightDistributionaluminBeforeAfterdeformation.pdf}
shows the surface roughness height distribution $P_h$ of a sandblasted aluminum surface 1 before (red line), and after (black line) squeezing the steel ball against the aluminum block. In the latter case the line scan data is from inside the indentation and obtained by first removing the macroscopic curvature. The sharp peak is due to the flat upper surfaces of the plastically deformed asperities.

The blue line in Fig.~\ref{1x.2height.37.65.63.full.pdf} 
shows the surface topography of the sandblasted surface 2 after indenting it with the ceramic ball. The linescan is from inside the indented region after removing the surface curvature. Similar result for the polished aluminum surface is shown by the upper green line. The lower green line is the measured surface topography of the polished aluminum surface before indenting it with the ball.

Figure~\ref{1x.2height.37.65.63.magnification.pdf} shows a magnified view of segments from the roughness profiles in Fig.~\ref{1x.2height.37.65.63.full.pdf} (the region between the two vertical lines). Note that the short-wavelength roughness in the plastically deformed region of the sandblasted surface (blue line), is very similar to that in the indented region of the polished surface (upper green line), and slightly larger than that of the original polished surface (green line). The short wavelength roughness is mainly due to the surface roughness of the ceramic ball (not shown), but in addition some contribution to the roughness may be due to inhomogeneous plastic flow.

Figure~\ref{1h.2Ph.37.29.pdf} 
shows the height probability distribution for the sandblasted surface 2 (original surface) (blue), and of from inside the indented region after removing the surface curvature (black). 

We have also performed indentation studies using a glass ball. The blue line in Fig.~\ref{1x.2h.with.glass.ball.pdf} 
shows the surface roughness height profile $h(x)$ of the sandblasted aluminum surface 2 after squeezing the silica glass ball against the aluminum block. The green line is the same result for the polished aluminum surface and the pink line is the topography of the glass ball. Note that the short wavelength roughness of both the sandblasted and polished aluminum surface are very similar to that of the glass ball.
This is due to the plastic imprint of the glass ball roughness in the contact regions with the aluminum surface. Recall that the penetration hardness of the silica glass is several times higher than that of the aluminum so only the aluminum will flow plastically. 

Figure~\ref{1logq.2logC.glassBall.sandblasted2.full.pdf} 
shows the surface roughness power spectrum as a function of the wavenumber (log-log scale) before indentation (blue), and after (black) plastic deformation. The results are for the sandblasted surface 2 indented by the glass ball.

Figure~\ref{1h.2Ph.sandblasted2.glassBall.pdf} 
shows the height probability distribution of the sandblasted surface 2 (blue line), and after indenting it with the glass ball (black curve). The height probability distribution of the indented surface is from inside the indented region after removing the surface curvature.

\begin{figure}
\includegraphics[width=0.95\columnwidth]{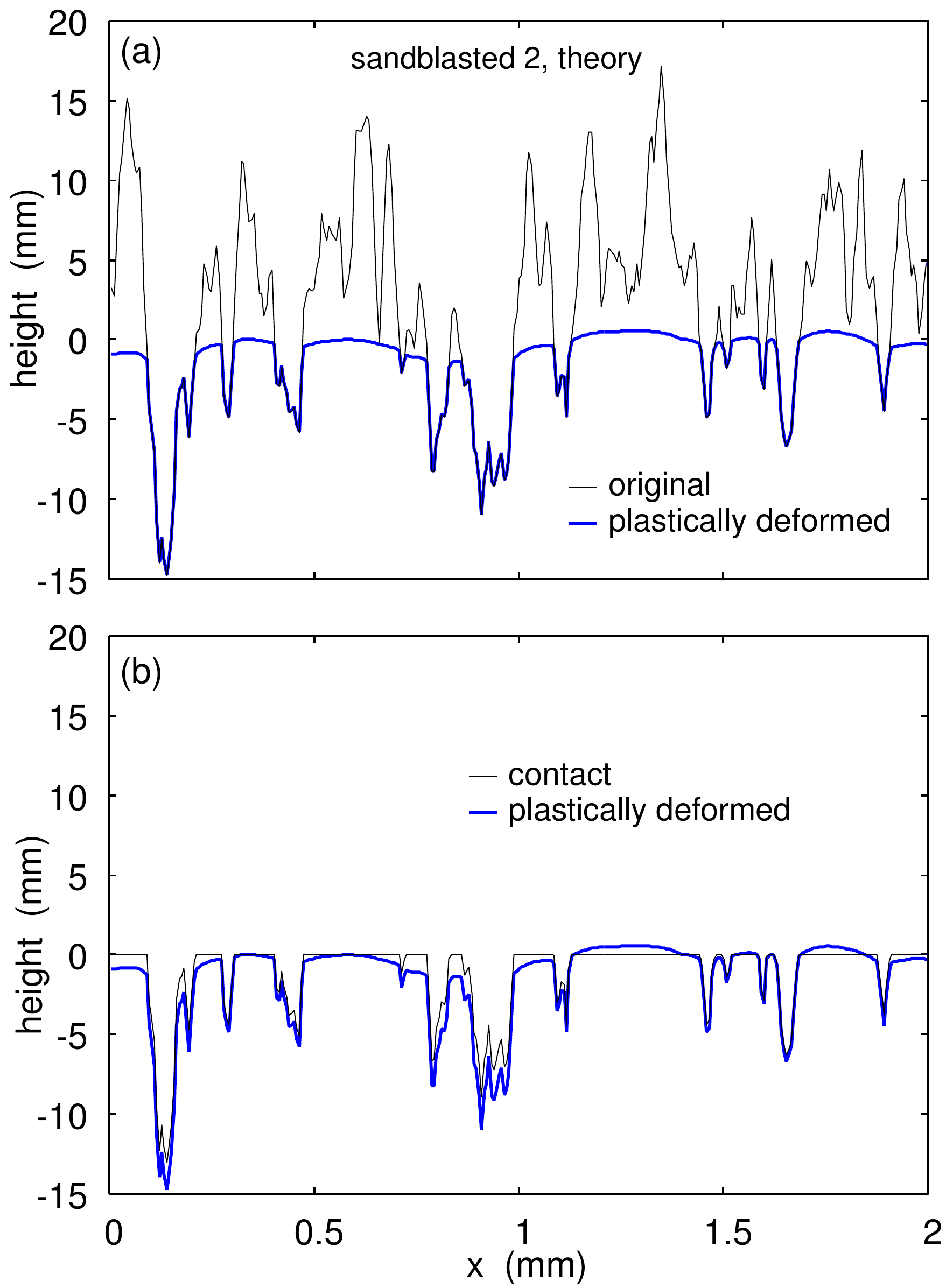}
\caption{\label{1x.2h.theory.plastic.and.original.and.contact.pdf}
Calculated plastic and elastoplastic deformations due to a load corresponding to the nominal contact pressure  $1 \ {\rm GPa}$. Depicted in (a) are the original surface topography (thin line), and the plastically deformed profile (thick line), and depicted in (b) are the surface topography during contact with a flat surface (thin line), and the plastically deformed profile (thick line).
}
\end{figure}

\begin{figure}
\includegraphics[width=0.95\columnwidth]{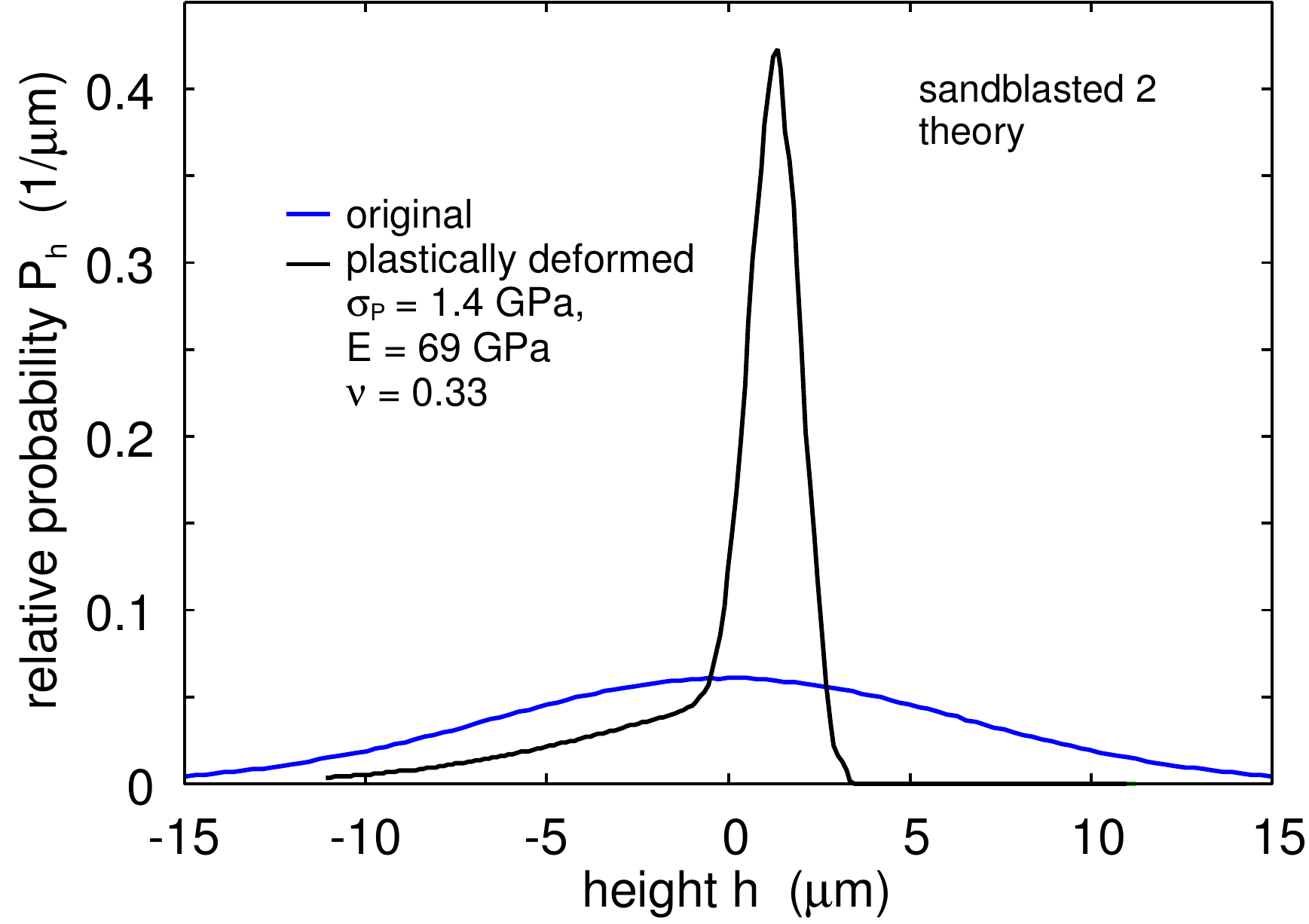}
\caption{\label{1h.2Ph.2048.theory.pdf}
The calculated height probability distribution
before (blue) and after (black) squeezing the rough surface against a perfectly smooth surface with
the nominal contact pressure $1 \ {\rm GPa}$.
}
\end{figure}

\begin{figure}
\includegraphics[width=0.95\columnwidth]{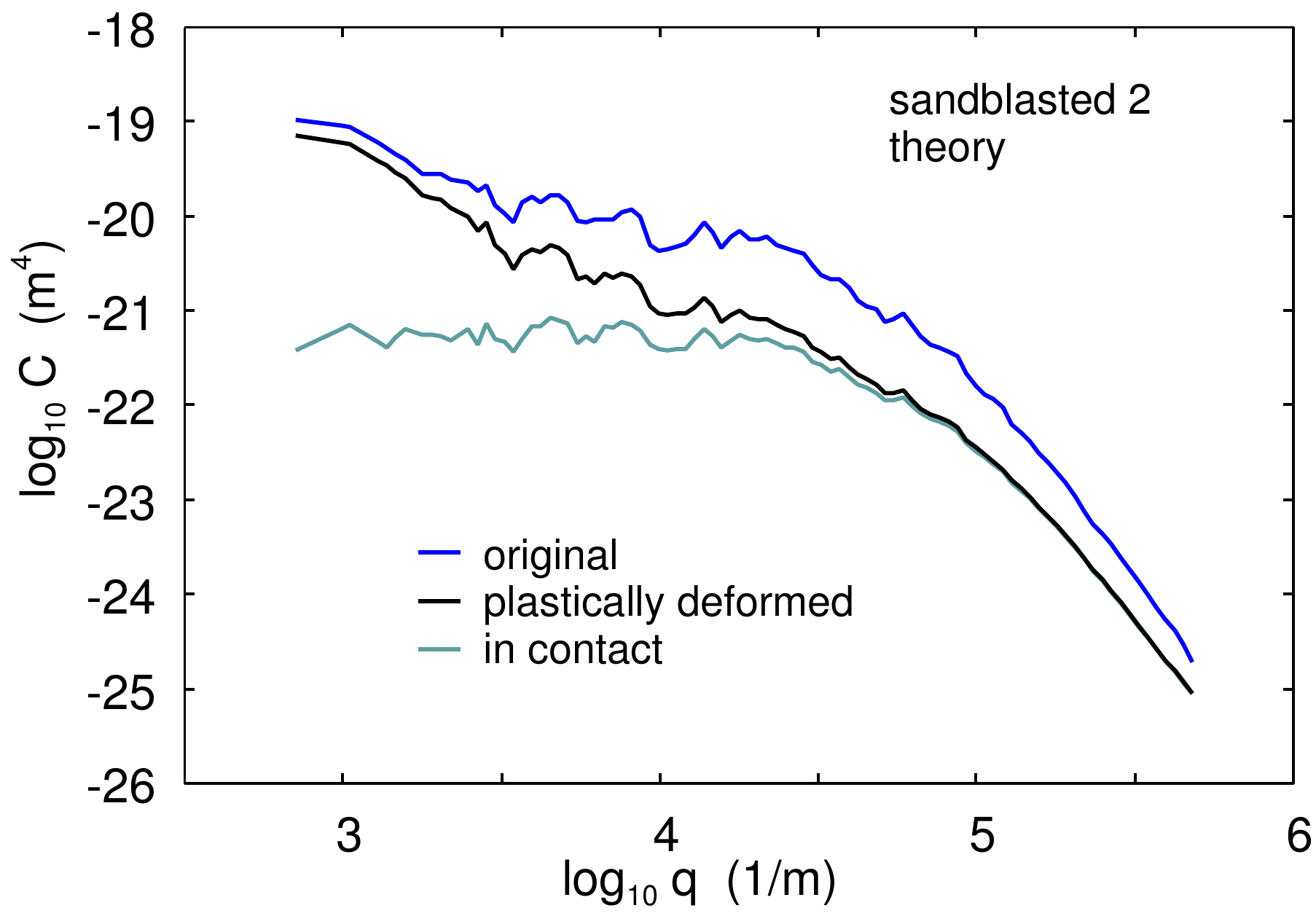}
\caption{\label{1logq.2logC2D.2048.theory.pdf}
The calculated surface roughness power spectrum as a function of the wavenumber (log-log scale)
before indentation (blue), and after (black) plastic deformation with
the nominal
contact pressure $1 \ {\rm GPa}$, and of the deformed surface
during indentation (contact pressure $1 \ {\rm GPa}$) (gray).
}
\end{figure}

\begin{figure}
\includegraphics[width=0.95\columnwidth]{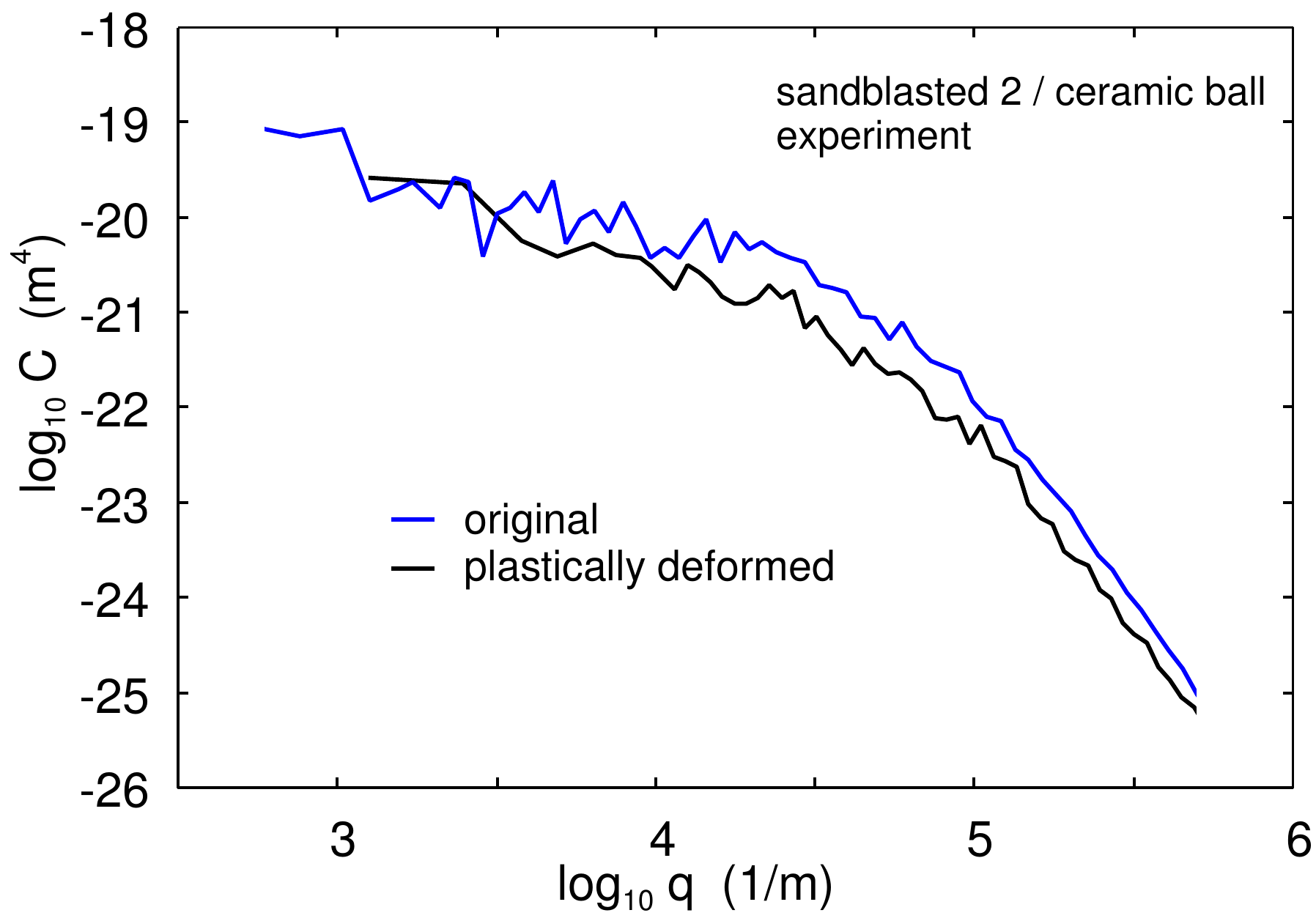}
\caption{\label{1logq.2logC.23.35.36.37.38.reduced.scale.pdf}
The measured surface roughness power spectrum (for the ceramic ball on the sandblasted surface 2)
as a function of the wavenumber (log-log scale)
before indentation (blue), and after (black) plastic deformation. The same as in Fig.
\ref{1logq.2logC.rough2.23.and.35-38.pdf} but plotted on a reduced wavenumber scale.
}
\end{figure}

\begin{figure}
\includegraphics[width=0.95\columnwidth]{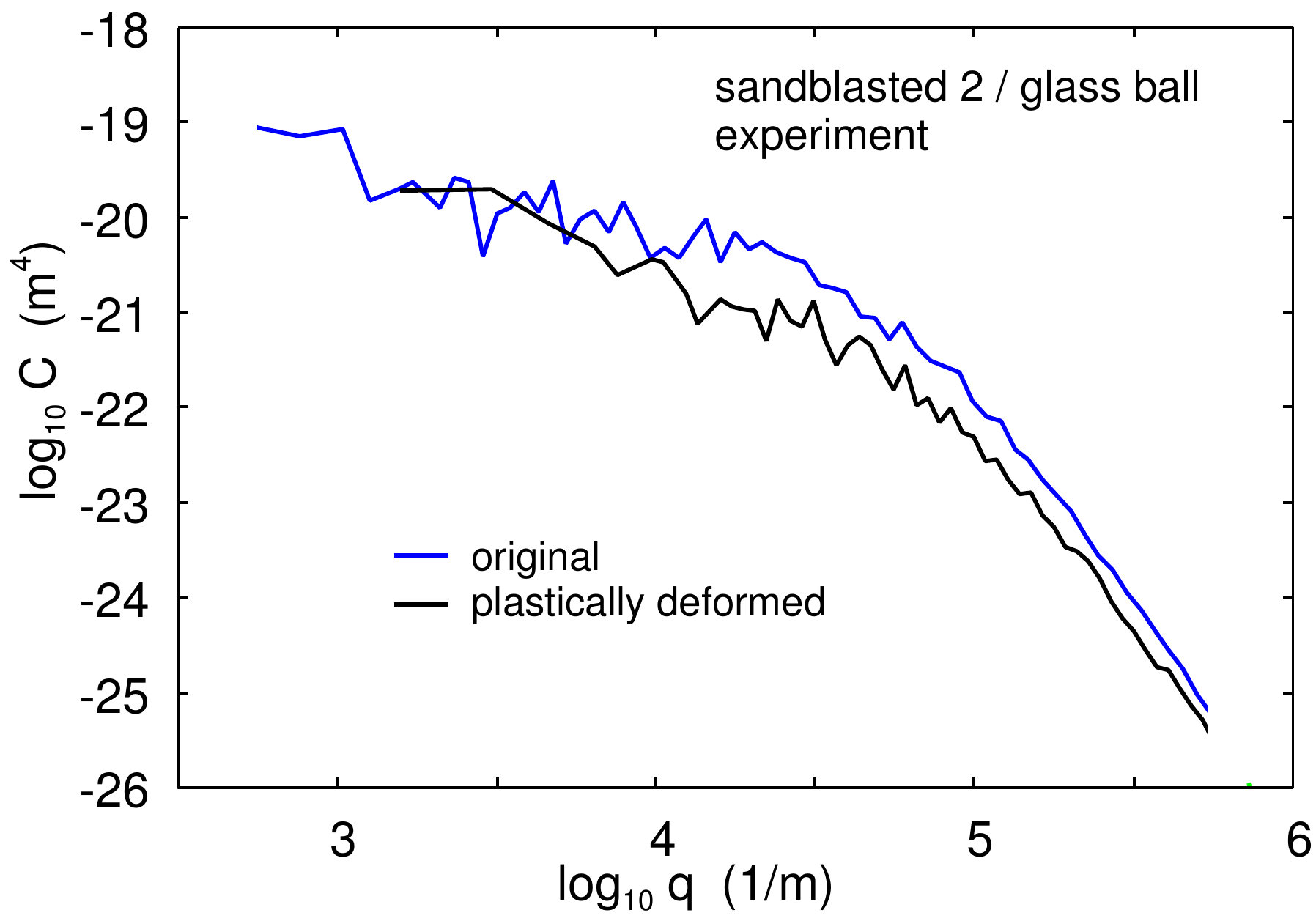}
\caption{\label{1logq.2logC.glassBall.sandblasted2.pdf}
The measured surface roughness power spectrum (for the glass ball on the sandblasted surface 2)
as a function of the wavenumber (log-log scale)
before indentation (blue), and after (black) plastic deformation. The same as in Fig.
\ref{1logq.2logC.glassBall.sandblasted2.full.pdf} but plotted on a reduced wavenumber scale.
}
\end{figure}

\begin{figure}
\includegraphics[width=0.95\columnwidth]{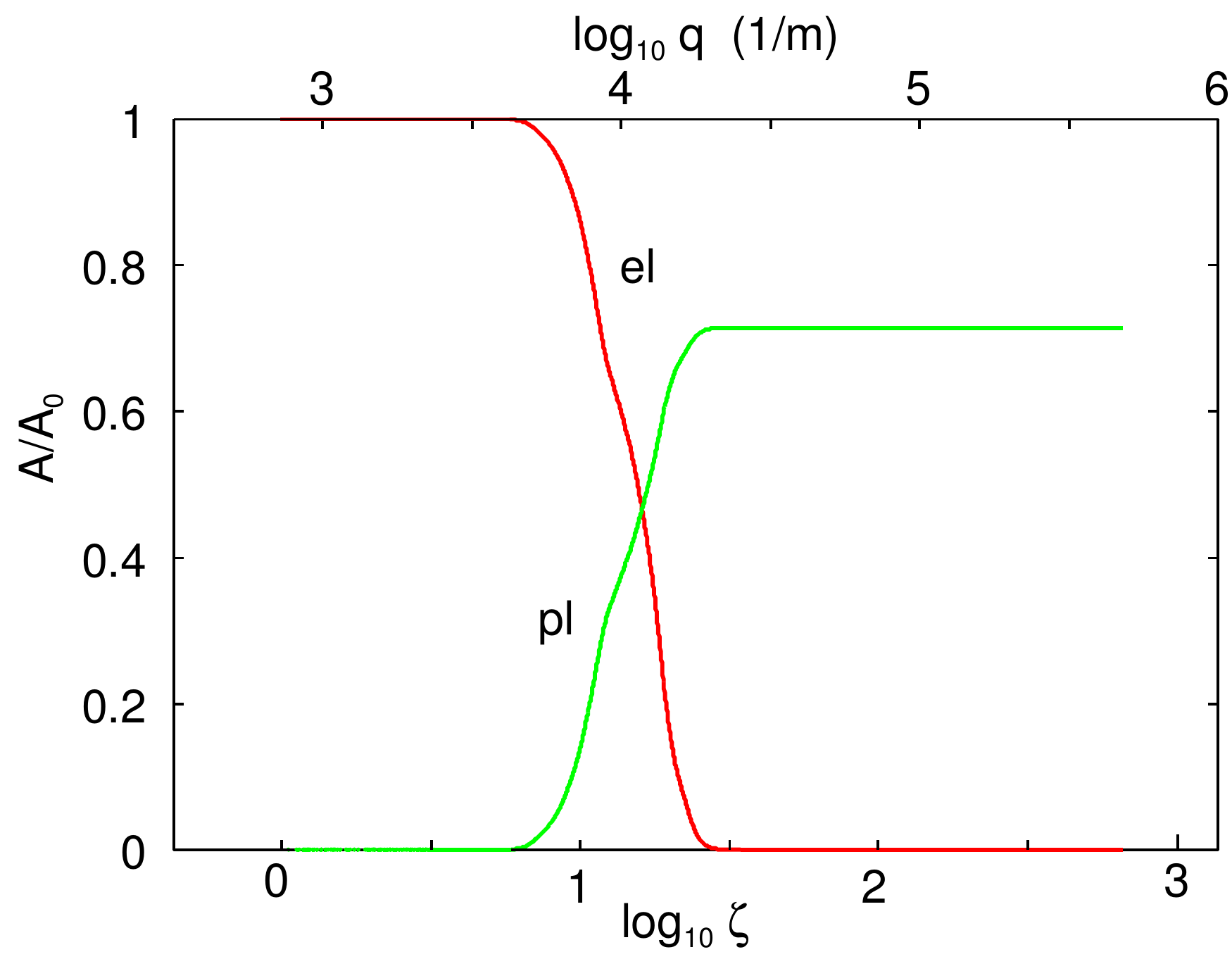}
\caption{\label{1logq.2Ael.Apl.pdf}
The relative contact area $A/A_0$ as a function of the magnification $\zeta$ (lower log-scale)
or as a function of the wavenumber $q=\zeta q_0$ (upper log-scale) as obtained using the Persson contact
mechanics theory with the power spectrum shown in Fig. \ref{1logq.2logC2D.2048.theory.pdf} (blue line).
The relative elastic contact area and plastic contact area
are shown separately as the red and green lines, respectively. In the calculation we
have used the elastoplastic parameters given in the text.
}
\end{figure}

\vskip 0.3cm
{\bf 4 Numerical simulation}

We have performed numerical simulations of the plastic deformation of rough surfaces.
We use the boundary element method (BEM) first described in \cite{AA1}, which treats the elastic deformations exactly (within the small slope approximation), and the plastic deformations is considered within the elastoplastic approximation. In this model the solid deform elastically as long as the surface stress is below the penetration hardness $\sigma_{\rm P}$. When the local stress reach the penetration hardness, the solid flow plastically without work hardening. In the model the plastic flow is taken into account by moving the surface grid point downwards in such a way that the local stress in the plastically deformed area is always equal to the penetration hardness. The numerical solution procedure is based on spectral theory, and an FFT-accelerated approach is applied to increase the computational efficiency. Inputs to the model is the geometries and roughness of the contacting bodies, their Young's modulus of elasticity, the Poisson ratios, and the indentation hardness of the softer of the two surfaces. It predicts the contact pressure distribution and the corresponding elastic and plastic deformations of the contacting bodies. Parameters like the real area of contact, the ratio of plastically deformed to the nominal contact area, can also be deduced by postprocessing the results. This BEM-based elastoplastic approach has also been frequently employed in other works, see e.g. \cite{AA2,Alm,AA4,AA5,AA6,AA7,AA8,AA9}.

We have performed calculations for the sandblasted surface 2. Since the topography of this surface was measured only along a line, we first calculated the 2D surface roughness power spectrum from the 1D power spectrum\cite{Nyak,CarbLor}. Next we generated mathematically a randomly rough surface with the powers spectrum of surface 2 using the procedure described in Appendix D in Ref. \cite{R1}. However, we have not used the full power spectrum as this would result in a surface with roughness over too many length scales or degrees of freedom. Thus, the surface we use has the size $2048 \times 2048$ grid points, and reproduce the measured power spectrum for $q < q_1$ with $q_1 \approx 5\times 10^5 \ {\rm m}^{-1}$.

We consider now squeezing a rigid and perfectly flat surface against an elastoplastic solid with the surface roughness obtained as described above. We assume the nominal contact pressure $1 \ {\rm GPa}$ which is similar as the nominal contact pressure acting in the indented region in the experiments. We assume the Young's elastic modulus $E=69 \ {\rm GPa}$, the Poisson ratio $\nu = 0.33$, which are typical values for aluminum. We also assume the aluminum penetration hardness $1.4 \ {\rm GPa}$.

The rough surface has the size $L\times L$ with $L\approx 12.5 \ {\rm mm}$. In Fig.~\ref{1x.2h.theory.plastic.and.original.and.contact.pdf} 
we show a $2 \ {\rm mm}$ long line scan of the the calculated surface topography. In (a) we show the original surface topography (thin line), and of the plastically deformed profile (thick line). In (b) we show the surface topography during contact with a flat surface (thin line), and of the plastically deformed profile (thick line). Note that in (b) the elastic rebound makes the upper surface of the plastically deformed asperities curved. Note also in (a) that the surface roughness below the plastically deformed region is unchanged. Since the total volume of the solid must be (nearly) unchanged, in reality material must flow also in the tangential direction, which result in some modification of the roughness also in the regions which was not in contact with the flat countersurface. For materials which undergoes work hardening this tangential flow becomes very important and can result in a complete breakdown of the plastic flow procedure used in the present paper, as indeed observed for polyethylene in Ref. \cite{Av} (see also Discussion). 

Figure~\ref{1h.2Ph.2048.theory.pdf} 
shows the calculated height probability distribution before (blue) and after (black) squeezing (and removing) the rough surface against the flat rigid countersurface. Both the original and plastically deformed surface have similar height distribution as observed in the experiment (compare to Fig.~\ref{1h.2Ph.37.29.pdf} and Fig.~\ref{1h.2Ph.sandblasted2.glassBall.pdf}). 

Figure~\ref{1logq.2logC2D.2048.theory.pdf} 
shows the calculated surface roughness power spectrum as a function of the wavenumber (log-log scale) before indentation (blue), and after plastic deformation (black). Also shown is the power spectrum of the surface of the solid when in contact with the flat rigid countersurface (gray). Note that for large wavenumber the plastically deformed surface and the surface in contact with the flat rigid surface exhibit the same surface roughness power spectrum. This is due to the fact that the large wavenumber roughness is due to the surface roughness in the regions not in contact with the flat surface, and this part is nearly unchanged. However, the surface area occupied by this (nearly unchanged) surface roughness is smaller than for the original surface, and this explain why the power spectrum for large wavenumbers of the plastically deformed surface is smaller than for the original surface.

In order to compare the measured power spectrum with the calculated ones, we show again 
in Fig.~\ref{1logq.2logC.23.35.36.37.38.reduced.scale.pdf} 
and Fig.~\ref{1logq.2logC.glassBall.sandblasted2.pdf} the measured power spectra 
(from Fig.~\ref{1logq.2logC.rough2.23.and.35-38.pdf} and Fig.~\ref{1logq.2logC.glassBall.sandblasted2.full.pdf})
but now on the same wavenumber interval as in Fig.~\ref{1logq.2logC2D.2048.theory.pdf}.
Comparing Fig.~\ref{1logq.2logC2D.2048.theory.pdf} with Fig.~\ref{1logq.2logC.23.35.36.37.38.reduced.scale.pdf} and \ref{1logq.2logC.glassBall.sandblasted2.pdf}, we conclude that the calculated power spectrum of the plastically deformed surface is similar to the measured one, but shows a slightly larger reduction in the magnitude of $C(q)$ then the measured one. This may indicate that the penetration hardness we used is slightly too small, but we have made no attempt to optimize the agreement with the experiments by changing the magnitude of $\sigma_{\rm P}$.

The Persson contact mechanics theory\cite{BP} can be used to study the nature of the contact area as we increase the magnification. When we study the interface at the magnification $\zeta$ we only observe surface roughness with wavenumber $q < \zeta q_0$, where $q_0$ is the smallest wavenumber. Thus for $\zeta = 1$ (or ${\rm log}_{10} \zeta = 0$) we do not observe any roughness and since the nominal contact pressure $p=1 \ {\rm GPa}$ is below the penetration hardness stress $\sigma_{\rm P} = 1.4  \ {\rm GPa}$
there is no plastic deformation, i.e. $A_{\rm el}/A_0 =1$ and $A_{\rm pl}/A_0 =0$.  When we increase the magnification we observe surface roughness and the contact area decreases and the contact stress increases until it becomes large enough to induce plastic flow.

Figure~\ref{1logq.2Ael.Apl.pdf} 
shows the relative contact area $A/A_0$ as a function of the magnification $\zeta$ (lower scale) or as a function of the wavenumber  $q=\zeta q_0$ (upper scale) as obtained using the Persson contact mechanics theory with the power spectrum shown in Fig.~\ref{1logq.2logC2D.2048.theory.pdf} (blue line). The relative elastic contact area and plastic contact area are shown separately as the red and green lines, respectively. In the calculation we have used the same  elastoplastic parameters as in the numerical simulations using the BEM-based approach. 

The result in Fig.~\ref{1logq.2Ael.Apl.pdf} are consistent with the power spectra shown in Fig.~\ref{1logq.2logC2D.2048.theory.pdf}-\ref{1logq.2logC.glassBall.sandblasted2.pdf}. Thus, Fig.~\ref{1logq.2Ael.Apl.pdf} shows that the long wavelength roughness components for ${\rm log}_{10}q < 3.5$ are elastically deformed, and this explain why in this wavenumber region the power spectrum of the 
plastically deformed surface is close to that of the original surface in Fig.~\ref{1logq.2logC2D.2048.theory.pdf}-\ref{1logq.2logC.glassBall.sandblasted2.pdf}.
For ${\rm log}_{10}q > 4.2$ the Persson theory predict that the contact is fully plastic which explain why the power spectrum of the plastically deformed surface is the same in and out of contact with the flat surface in this wavenumber region.

\begin{figure}
\includegraphics[width=0.95\columnwidth]{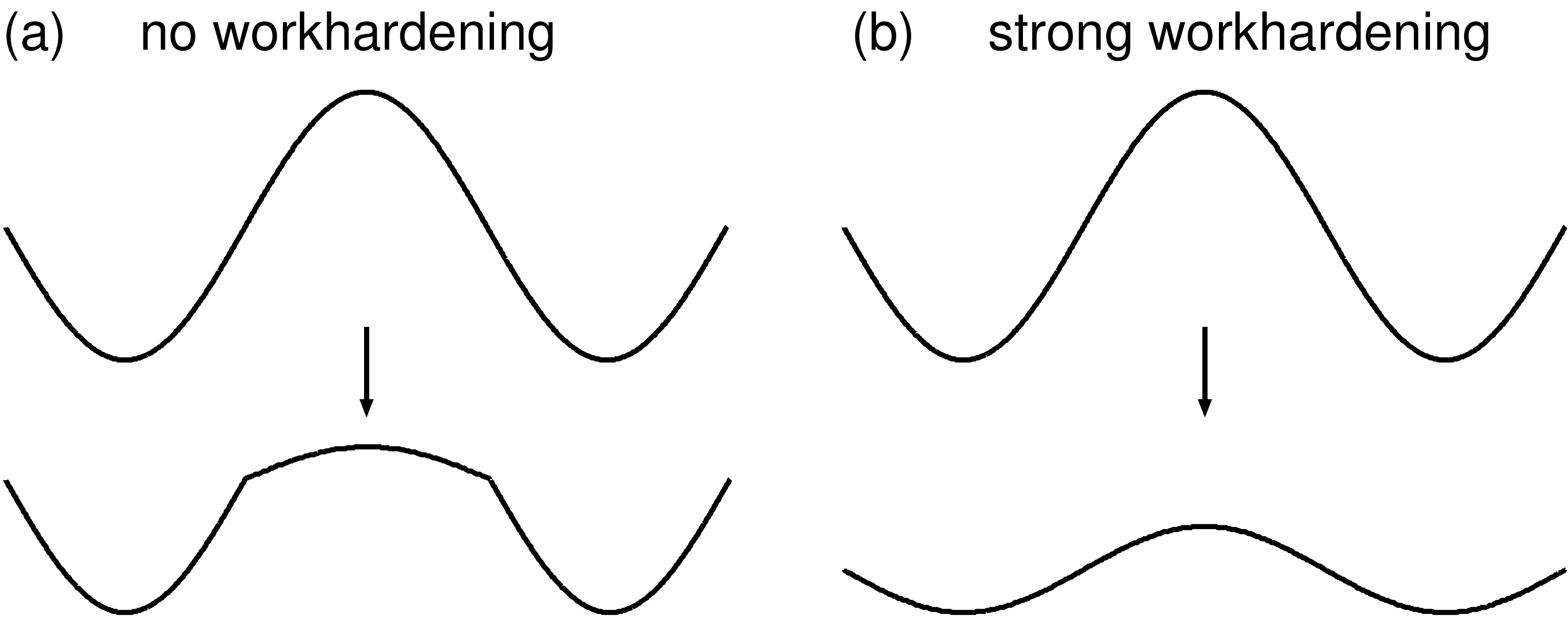}
\caption{\label{1x.2h.two.ways.pdf}
Two limiting behavior of the plastic deformation of a surface sinus-roughness-component.
(a) For a material without work hardening (e.g., a material already strongly work hardened)
the plastic flow is localized to the region where the solid make contact with the countersurface,
here assumed to be a flat rigid surface. (b) For a material which undergoes strong work hardening the
flow of material is more long range, resulting in material transfer to the valley region,
effectively resulting in a sinus profile with smaller amplitude.
}
\end{figure}

\vskip 0.3cm
{\bf 5 Discussion}

The simple procedure used above to describe the plastic flow produces plastically deformed surfaces with height probability distributions and surface roughness power spectra in semi-quantitative agreement with the experimental data. In our study we used an aluminum block, but the results are probably applicable for other metals as well, assuming negligible work hardening so the simple elastoplastic description with a constant penetration hardness is a reasonable approximation. However, the following points need to be taken into consideration:

(a) The penetration hardness depends on the length scale. Suppose we indent a solid with a rigid perfectly smooth sphere. If we look at the indentation at low magnification we do not see any surface roughness and we would calculate the penetration hardness $\sigma_{\rm P} = F_{\rm N}/A_0$, where $A_0=\pi r_0^2$ is the projected contact area (a circular area). This is, in fact, the standard definition of the penetration hardness. However, in general we do not make plastic contact everywhere within the apparent (projected) contact area $A_0$. It is clear from this fact that the penetration hardness at the asperity length scale must be higher than at the macroscopic length scale. If we increase the magnification further we may observe that within the plastically deformed (macro) asperity contact regions there may be regions which are not plastically deformed, corresponding to an even higher penetration hardness at even shorter length scale. To obtain the correct contact mechanics observed at high magnification is is necessary to include the length (or magnification) dependency of the penetration hardness.

(b) At very short length scale the plastic flow may be inhomogeneous. This implies that if one indent a perfectly smooth surface of a solid with a spherical ball with perfectly smooth surface,  roughness may be generated in the indented surface area\cite{Pas}

(c) The procedure used above to describe the plastic flow gives plastically deformed surfaces with roughness in relative good agreement with experiment for the aluminum block we used. But this result is expected only if there is no work hardening. The aluminum block we used has probably already undergone strong work hardening during the preparation process, and the results presented above may not be valid for a thermally annealed metal block.  In an earlier study we have found that for some polymers, in particular polyethylene, the plastically deformed surface exhibit a perfectly symmetric Gaussian-like height probability distribution, in contrast to the strongly skewed height distribution we observe for aluminum (see Fig.~\ref{1h.2Ph.37.29.pdf} and Fig.~\ref{1h.2Ph.2048.theory.pdf}). We interpret this as resulting from strong work hardening, which effectively result in flow of materials, from the top of asperities to the nearby wells as indicated in Fig.~\ref{1x.2h.two.ways.pdf}. 
Only by assuming this can the experimental data for polymers be understood.

The study presented in this paper is relevant for the fluid leakage in metallic seals. Metallic seals are usually made from steel, copper or bronze, and these metals usually have work hardened surface layers and should deform plastically in a similar way as the aluminum 
block studied here. Thus, we believe that the theory approach used here in combination with, e.g. the critical junction theory, may be used to estimate the leakage of metallic seals. Such an (experimental and theory) study will be reported on elsewhere\cite{Fi}.

\vskip 0.3cm
{\bf 6 Summary and conclusion}

We have presented experimental and matching computer simulation results pertaining to indentation of rough aluminum surfaces with balls made of steel, ceramic and silica glass. We found that the BEM-based approach, with a simple way to include plasticity within the elastoplastic model description, can be used to predict the height probability distribution and the surface roughness power spectra of the plastically deformed surfaces. The experimental- and the numerical simulation results, are also consistent with the predictions of the Persson contact mechanics theory for the elastic and plastic contact area as a function of the magnification. The contact area is of direct importance for the leakage of seals as it determines when the contact area percolate and the leakage vanish. This study, combined with findings in \cite{Av}, indicate that work hardening will strongly affect the nature of the plastically deformed roughness: for material with strong work hardening there is, at present, no simple numerical approach to predict the surface topography of plastically deformed surfaces.

\vskip 1.5cm
{\it Acknowledgments:}
BNJP acknowledges support by the DFG-grant: PE 807/12-1.\\
\noindent AA acknowledgeds support from VR (The Swedish Reseach Council): DNR 2019-04293.

\end{document}